\documentclass[3p,preprint]{elsarticle}

\usepackage{amssymb, amsfonts, latexsym, mathtools} %
\usepackage{graphicx, setspace, subfigure}
\usepackage{amsmath,euscript,mathrsfs}
\usepackage[hidelinks]{hyperref} 
\usepackage[usenames,dvipsnames,svgnames]{xcolor}

\hypersetup{
    colorlinks,
    linkcolor={blue!50!black},
    citecolor={blue!50!black},
    urlcolor={blue!50!black}
}

\graphicspath{{../Figures/}}

\makeatletter
\def\blfootnote{\gdef\@thefnmark{}\@footnotetext}
\makeatother

\usepackage[all]{hypcap}

\usepackage{array}
\newcolumntype{C}[1]{>{\centering\arraybackslash}m{#1}}

\usepackage[super]{nth}

\usepackage{comment}

\DeclarePairedDelimiter{\abs}{\lvert}{\rvert}

\allowdisplaybreaks[1]

\newcommand{\lb}{\left (}
\newcommand{\rb}{\right )}

\newcommand{\lsq}{\left [}
\newcommand{\rsq}{\right ]}

\renewcommand{\O}[1]{O \lb #1 \rb}

\newcommand{\eqtext}[1]{\quad \text{#1} \quad}
\newcommand{\RA}{\quad \Rightarrow \quad}

\newcommand{\sechn}[2]{\mathop{\rm sech}\nolimits ^{#1} \lb #2 \rb}

\newcommand{\arccosh}[1]{\mathop{\rm arccosh}\nolimits \lb #1 \rb}

\journal{Wave Motion}

\begin{document}

\begin{frontmatter}

\title{Solitary Wave Propagation in Elastic Bars with Multiple Sections and Layers} %
\author[lbo]{M. R. Tranter\corref{cor1}}
\ead{M.R.Tranter@lboro.ac.uk}
\cortext[cor1]{Corresponding author.}
\address[lbo]{Department of Mathematical Sciences, Loughborough University, Loughborough LE11 3TU, United Kingdom}


\begin{abstract}%
In this paper we present a numerical scheme for solving a system of Boussinesq-type equations. This can correspond to longitudinal displacements in a multi-layered elastic bar with delamination, with conditions on the interface between the sections of the bar. The method is initially presented for two coupled equations in each section and multiple sections in the bar, and later extended to any number of layers. Previous works have presented a similar method constructed using finite-difference methods, however these only solved for two sections of the bar at a time which limited the scope of studies using these methods. 

The new method presented here solves for all sections at a given time step and therefore the transmitted and reflected waves in each section of the bar can be studied. The new results are shown to be in excellent agreement with previously obtained results and a further study is performed showing that the delamination width can be inferred from the changes to the incident soliton. The generalised form of this method, for any number of sections and layers with coupling terms independent of time derivatives, can be used to study the behaviour of longitudinal waves in more complicated waveguides in future studies.
\end{abstract}

\begin{keyword}
finite-difference scheme \sep solitary wave \sep delamination
\end{keyword}

\end{frontmatter}

\section{Introduction}
\label{sec:Intro}
Many recent developments in the field of solid mechanics have been dedicated to the description of long longitudinal bulk strain solitary waves in elastic waveguides, for example in rods and plates \cite{Samsonov01, Porubov03}. The existence of longitudinal bulk strain solitons, as predicted by this theory, was confirmed in experiments \cite{Dreiden88, Samsonov98, Semenova05}. The theoretical and experimental studies were more recently extended to the case of adhesively bonded layered bars, with delamination \cite{Khusnutdinova08, Khusnutdinova09, Dreiden08, Dreiden11, Dreiden12, Dreiden14a}. These studies showed that longitudinal bulk waves in a layered bar with delamination can be described by a system of Boussinesq-type equations, with continuity of longitudinal displacement and normal stress at the interface between sections. This system can either be formed of uncoupled (perfect bonding) or coupled Boussinesq equations (soft bonding layer) \cite{Khusnutdinova08, Khusnutdinova09}.

In the case of coupled regularised Boussinesq (cRB) equations, the soliton solution is replaced by a \textit{radiating solitary wave}, that is a solitary wave with a one-sided, co-propagating oscillatory tail \cite{Khusnutdinova09}. The study of the initial value problem for the cRB equations has shown the emergence of such radiating solitary waves for closely matched characteristic speeds \cite{Khusnutdinova11} and an analytical description of these waves has recently been constructed \cite{Grimshaw17}. Radiating solitary waves have also been observed experimentally in layered bars with a soft bonding layer \cite{Dreiden11, Dreiden12}.

Finite-difference techniques are commonly used to solve PDEs \cite{Gerald84, Smith85}, but they can also be extended to more complicated equations such as the Korteweg-de Vries (KdV) equation \cite{Drazin89, Marchant02}, the Kadomtsev-Petviashvili equation \cite{Bratsos98}, Boussinesq equation or cRB equations \cite{Khusnutdinova11}, to name a few. Our recent work involved the development of a finite-difference approach to solve the wave scattering problem for Boussinesq-type and cRB equations, and describe the transmission and reflection of strain solitary waves that are generated \cite{Khusnutdinova15, Khusnutdinova17}. These methods were limited by the requirement that the developed numerical scheme can only solve for two sections of the bar at a time, resulting in a moving calculation window. The aim of this paper is to describe an extension to the numerical method outlined in \cite{Khusnutdinova15, Khusnutdinova17}, for the case when there are three or more sections in the bar. This extension removes the requirement that sections have to be of sufficient length so that the solution can be fully contained in a single section at a given time. 

The paper is structured as follows. In Section \ref{sec:DNM3SeccRB} we overview the previous approach to solving such a system of equations and outline the new method for a bar with $M$ sections and a soft bond between the layers i.e. the longitudinal displacements are described by cRB equations in the bonded sections and uncoupled Boussinesq equations in the delaminated sections. This can be adjusted to the case of a perfectly bonded bar by appropriately modifying coefficients (see \cite{Khusnutdinova08, Khusnutdinova15}). In Section \ref{sec:Results} we test the method for various configurations of the layered structure and compare the result of these calculations to previously obtained results for comparison. This approach is then compared to the results for a finite delamination in \cite{Khusnutdinova17} where only the semi-analytical method was used due to the previous limitations of the direct numerical method. The case of a finite delamination in a perfectly bonded bar is also studied. In Section \ref{sec:Conclusions} we conclude our discussions.

\section{Numerical Method for a Bonded Bar With \texorpdfstring{$M$}{M} Sections}
\label{sec:DNM3SeccRB}
\begin{figure}[ht]
	\begin{center}
		\includegraphics[width=0.5\linewidth, trim = 0mm 0mm 0mm 0mm]{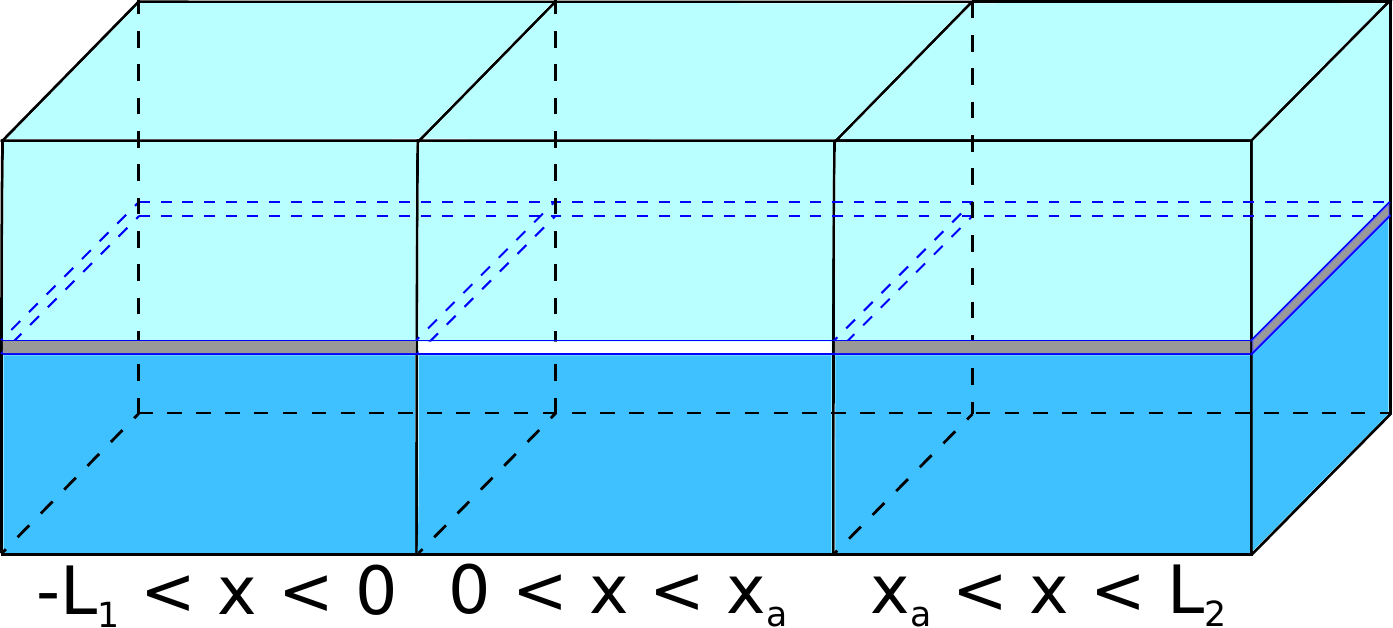}
		\caption{Two-layered bar with three sections, including a finite delamination at $0 < x < x_a$.}
		\label{fig:DelamBar3S}
	\end{center}
\end{figure}

A model for the longitudinal displacement in a layered bar with a soft bond between the layers was derived in \cite{Khusnutdinova09}, where it was shown that the displacements are governed by cRB equations. This configuration was studied in \cite{Khusnutdinova17, Khusnutdinova17a} for a semi-infinite or finite delamination, using the numerical method outlined in \cite{Khusnutdinova15}. The numerical scheme derived in \cite{Khusnutdinova15} and used in \cite{Khusnutdinova17} is applicable to two sections of a layered structure, however it can be used to find the solution for a bar of three or more sections, as done in these papers, with some conditions. Let us consider a bar with three sections, such as the one in Figure \ref{fig:DelamBar3S}, where the domains are $\Omega_1 = [-L_1, 0]$, $\Omega_2 = [0, x_a]$ and $\Omega_3 = [x_a, L_2]$. Note that the longitudinal direction is denoted by $x$.

In \cite{Khusnutdinova17} we initially computed the solution in $\Omega_1$ and $\Omega_2$ with two constraints: that the solitary wave has not reached the boundary between the second and third sections at $x_a$, and that the waves reflected from the interface at $x = 0$ have not reflected back from $x=-L_1$ and entered the second domain again at $x = 0$. A similar approach is then followed for the solution in $\Omega_2$ and $\Omega_3$, where we choose an appropriate time so that the waves in $\Omega_2$ have not reflected from the boundary $x = x_a$ and reached the boundary at $x = 0$. 

The method therefore has a limitation; if the solution is not localised, then we cannot compute the solution for all three sections without losing some information from the solution. In the case of a short finite delamination, the solution is unlikely to be fully contained in this section at any time and therefore some information will be lost when the calculation window is moved. We aim to rectify this restriction.

\subsection{Two-Layered Bar}
\label{sec:cRB}
Let us consider the system of cRB equations defined in \cite{Khusnutdinova17}, modelling a two-layered bar with a soft bond between the layers. In this case we have either cRB equations (in bonded sections) or uncoupled Boussinesq equations (in delaminated areas) in each section. To simplify the notation, we will assume that we have cRB equations in each section. We note that the matching at the interfaces is the same regardless of the equation type and therefore this generalisation is valid. This allows us to write the equations in each section in the following convenient form:
\begin{align}
u_{tt}^{(l)} - u_{xx}^{(l)} &= 2 \epsilon \lsq -6 u_{x}^{(l)} u_{xx}^{(l)} + u_{ttxx}^{(l)} - \delta_{l} \lb u^{(l)} - w^{(l)} \rb \rsq, \notag \\
w_{tt}^{(l)} - c_{l}^2 w_{xx}^{(l)} &= 2 \epsilon \lsq -6 \alpha_{l} w_{x}^{(l)} w_{xx}^{(l)} + \beta_{l} w_{ttxx}^{(l)} + \gamma_{l} \lb u^{(l)} - w^{(l)} \rb \rsq, \quad x_{l-1} < x < x_{l}, \label{DNMcRBEq} 
\end{align}
where $u$ and $w$ are the longitudinal displacements in the upper and lower layers respectively, $x_{l}$ is the position of the interface between the sections, and we have $l = 1, \dots, M$ sections. We have the coefficients $\alpha_{l}$, $\beta_{l}$, $c_{l}$ and coupling constant $\gamma_{l}$ in the lower layers, and we have assumed that the only unknown constant in the upper layer is the coupling constant $\delta_{l}$, which can be achieved through a rescaling of parameters. The scheme could be written for general coefficients in each layer however this would result in a more complicated notation being required. 

Here we have assumed that the coefficients $\alpha_{l}, \beta_{l}, c_{l}$ can differ between sections. Therefore we introduce the subscript notation for coefficients, identifying the section where each coefficient is applied. The small wave parameter, $\epsilon$, is the same in all sections. We have continuity of longitudinal displacement,
\begin{align}
u^{(l)} |_{x=x_{l}} &= u^{(l+1)} |_{x=x_{l}}, & w^{(l)} |_{x=x_l} &= w^{(l+1)} |_{x=x_l}, &l=1,\dots,M-1, \label{DNMcRBC1}
\end{align}
and continuity of normal stress at the interface between the sections,
\begin{align}
\left. u_{x}^{(l)} + 2\epsilon \lsq -3 \lb u_{x}^{(l)} \rb^2 + u_{ttx}^{(l)} \rsq \right |_{x=x_l} &= \left. u_{x}^{(l+1)} + 2 \epsilon \lsq -3 \lb u_{x}^{(l+1)} \rb^2 + u_{ttx}^{(l+1)} \rsq \right |_{x=x_l},  \notag \\
\left. c_l^2 w_{x}^{(l)} + 2\epsilon \lsq -3 \alpha_l \lb w_{x}^{(l)} \rb^2 + \beta_l w_{ttx}^{(l)} \rsq \right |_{x=x_l} &= \left. c_{l+1}^2 w_{x}^{(l+1)} + 2\epsilon \lsq -3 \alpha_{l+1} \lb w_{x}^{(l+1)} \rb^2 + \beta_{l+1} w_{ttx}^{(l+1)} \rsq \right |_{x=x_l},  \label{DNMcRBC2}
\end{align}
for $l = 1, \dots, M - 1$. To solve the equation system \eqref{DNMcRBEq} - \eqref{DNMcRBC2}, we make use of central difference approximations in both space and time. Let us first discretise each domain $\lsq x_{l-1}, x_{l} \rsq \times \lsq 0, T \rsq$ into a grid with spatial discretisation $h_{l} = \Delta x$ and temporal discretisation $\kappa = \Delta t$. Note here that the spatial discretisation $h_{l}$ can differ across sections of the bar, allowing for an adaptive method to be applied so that a complex solution in one section can be computed with a finer mesh and a sparse mesh can be used when the solution is simpler. The number of points in each domain is given by $N_{l} = \lb x_{l} - x_{l-1} \rb/h_l$ and we have $i = 0, 1, \dots, N_{l}$ in each section.  The analytical solution $u^{(l)} \lb x, t \rb$ is approximated by the exact solution of the finite-difference scheme $u^{(l)} \lb i h_l, j \kappa \rb$, denoted $u_{i,j}^{(l)}$. Substituting the central difference approximations into \eqref{DNMcRBEq} gives a system of tridiagonal equations, of the form
\begin{align}
&-2 \epsilon u_{i+1,j+1}^{(l)} + \lb 4 \epsilon + h_l^2 \rb u_{i,j+1}^{(l)} - 2 \epsilon u_{i-1,j+1}^{(l)} = \lb \kappa^2 - 4\epsilon \rb \lb u_{i+1,j}^{(l)} - 2 u_{i,j}^{(l)} + u_{i-1,j}^{(l)} \rb + 2h_l^2 u_{i,j}^{(l)} \notag \\
&- \frac{6 \epsilon \kappa^2}{h_l} \lsq \lb u_{i+1,j}^{(l)} \rb^2 - \lb u_{i-1,j}^{(l)} \rb^2 - 2u_{i+1,j}^{(l)} u_{i,j}^{(l)} + 2u_{i,j}^{(l)} u_{i-1,j}^{(l)} \rsq - 2 \epsilon \delta_{l} h_l^2 \kappa^2 \lb u_{i,j}^{(l)} - w_{i,j}^{(l)} \rb \notag \\
&+ 2 \epsilon u_{i+1,j-1}^{(l)} - \lb 4\epsilon + h_l^2 \rb u_{i,j-1}^{(l)} + 2 \epsilon u_{i-1,j-1}^{(l)},
\label{DNMcRBDisc1}
\end{align}
and
\begin{align}
&- 2\epsilon \beta_l w_{i+1,j+1}^{(l)} + \lb 4\epsilon\beta_l + h_l^2 \rb w_{i,j+1}^{(l)} - 2\epsilon\beta_l w_{i-1,j+1}^{(l)} = \lb \kappa^2 c_l^2 - 4\epsilon\beta_l \rb \lb w_{i+1,j}^{(l)} - 2 w_{i,j}^{(l)} + w_{i-1,j}^{(l)} \rb + 2h_l^2 w_{i,j}^{(l)} \notag \\
& - \frac{6\epsilon \alpha_l \kappa^2}{h_l} \lsq \lb w_{i+1,j}^{(l)} \rb^2 - \lb w_{i-1,j}^{(l)} \rb^2 - 2 w_{i+1,j}^{(l)} w_{i,j}^{(l)} + 2 w_{i,j}^{(l)} w_{i-1,j}^{(l)} \rsq + 2 \epsilon \gamma_{l} h_{l}^2 \kappa^2 \lb u_{i,j}^{(l)} - w_{i,j}^{(l)} \rb \notag \\
&+  2\epsilon\beta_l w_{i+1,j-1}^{(l)} - \lb  4\epsilon\beta_l + h_l^2 \rb w_{i,j-1}^{(l)} +  2\epsilon\beta_l w_{i-1,j-1}^{(l)}, \quad l=1,\dots,M, 
\label{DNMcRBDisc2}
\end{align}
for $u_{i,j}$ and $w_{i,j}$ respectively. The condition for continuity of longitudinal displacement \eqref{DNMcRBC1} becomes
\begin{equation}
u_{N_l,j+1}^{(l)} = u_{0,j+1}^{(l+1)} \eqtext{and} w_{N_l,j+1}^{(l)} = w_{0,j+1}^{(l+1)}, \quad l = 1, \dots, M - 1.
\label{DNMcRBC1Disc}
\end{equation}
In continuity condition \eqref{DNMcRBC2} we make use of the central difference approximations and introduce ``ghost points'' of the form $u_{N_l+1,j+1}^{(l)}$, $u_{-1,j+1}^{(l+1)}$, $w_{N_l+1,j+1}^{(l)}$ and $w_{-1,j+1}^{(l+1)}$, so \eqref{DNMcRBC2} becomes
\begin{align}
&h_l \kappa^2 \lb u_{N_l+1,j+1}^{(l)} - u_{N_l-1,j+1}^{(l)} \rb -3 \kappa^2 \epsilon \lsq \lb u_{N_l+1,j+1}^{(l)} \rb^2 + \lb u_{N_l-1,j+1}^{(l)} \rb^2 - 2 u_{N_l+1,j+1}^{(l)} u_{N_l-1,j+1}^{(l)} \rsq \notag \\
&+ 2 h_l \epsilon \lb 2 u_{N_l+1,j+1}^{(l)} - 2 u_{N_l-1,j+1}^{(l)} - 5 u_{N_l+1,j}^{(l)} + 5 u_{N_l-1,j}^{(l)} \right. \notag \\
&\left. \hspace{0.399\linewidth} ~+ 4 u_{N_l+1,j-1}^{(l)} - 4 u_{N_l-1,j-1}^{(l)} - u_{N_l+1,j-2}^{(l)} + u_{N_l-1,j-2}^{(l)} \rb  \notag \\
=&~ \frac{h_{l}^{2}}{h_{l+1}} \kappa^2 \lb u_{1,j+1}^{(l+1)} - u_{-1,j+1}^{(l+1)} \rb - 3 \frac{h_{l}^2}{h_{l+1}^2} \kappa^2 \epsilon \lsq \lb u_{1,j+1}^{(l+1)} \rb^2 + \lb u_{-1,j+1}^{(l+1)} \rb^2 - 2 u_{1,j+1}^{(l+1)} u_{-1,j+1}^{(l+1)} \rsq \notag \\
&~+ 2 \frac{h_{l}^2}{h_{l+1}} \epsilon \lb 2 u_{1,j+1}^{(l+1)} - 2 u_{-1,j+1}^{(l+1)} - 5 u_{1,j}^{(l+1)} + 5 u_{-1,j}^{(l+1)} + 4 u_{1,j-1}^{(l+1)} - 4 u_{-1,j-1}^{(l+1)} -  u_{1,j-2}^{(l+1)} + u_{-1,j-2}^{(l+1)} \rb,
\label{DNMcRBC2Disc1}
\end{align}
and
\begin{align}
&h_l \kappa^2 c_l^2 \lb w_{N_l+1,j+1}^{(l)} - w_{N_l-1,j+1}^{(l)} \rb -3 \kappa^2 \epsilon \alpha_l \lsq \lb w_{N_l+1,j+1}^{(l)} \rb^2 + \lb w_{N_l-1,j+1}^{(l)} \rb^2 - 2 w_{N_l+1,j+1}^{(l)} w_{N_l-1,j+1}^{(l)} \rsq \notag \\
&~+ 2 h_l \epsilon \beta_l \lb 2 w_{N_l+1,j+1}^{(l)} - 2 w_{N_l-1,j+1}^{(l)} - 5 w_{N_l+1,j}^{(l)} + 5 w_{N_l-1,j}^{(l)} \right. \notag \\
&\left. \hspace{0.434\linewidth} ~+ 4 w_{N_l+1,j-1}^{(l)} - 4 w_{N_l-1,j-1}^{(l)} - w_{N_l+1,j-2}^{(l)} + w_{N_l-1,j-2}^{(l)} \rb \notag \\
=&~\frac{h_{l}^{2}}{h_{l+1}} \kappa^2 c_{l+1}^2 \lb w_{1,j+1}^{(l+1)} - w_{-1,j+1}^{(l+1)} \rb - 3 \frac{h_{l}^2}{h_{l+1}^2} \kappa^2 \epsilon \alpha_{l+1} \lsq \lb w_{1,j+1}^{(l+1)} \rb^2 + \lb w_{-1,j+1}^{(l+1)} \rb^2 - 2 w_{1,j+1}^{(l+1)} w_{-1,j+1}^{(l+1)} \rsq \notag \\
&~+ 2 \frac{h_{l}^2}{h_{l+1}} \epsilon \beta_{l+1} \lb 2 w_{1,j+1}^{(l+1)} - 2 w_{-1,j+1}^{(l+1)} - 5 w_{1,j}^{(l+1)} + 5 w_{-1,j}^{(l+1)} + 4 w_{1,j-1}^{(l+1)} - 4 w_{-1,j-1}^{(l+1)} -  w_{1,j-2}^{(l+1)} + w_{-1,j-2}^{(l+1)} \rb,
\label{DNMcRBC2Disc2}
\end{align}
for $u$ and $w$ respectively, and for $l = 1, \dots, M - 1$. We assume that the bar is not deformed at both ends so that we can enforce zero strain, i.e. $u_{x} = 0$ at $x = x_{0}$ and $x = x_{M}$, and similarly for $w$. Applying a central difference approximation to this condition, we have
\begin{equation}
u_{1,j+1}^{(1)} = u_{-1,j+1}^{(1)} \eqtext{and similarly} u_{N_{M}+1,j+1}^{(M)} = u_{N_{M}-1,j+1}^{(M)},
\label{DNMcRBBCDisc}
\end{equation}
and equivalent relations for $w$. This numerical method is an implicit, fully-discrete finite-difference scheme and, as second-order finite-difference approximations are used everywhere, the scheme is second-order in both space and time. The truncation error and stability of the finite-difference scheme for the coupled Boussinesq equations was discussed in \cite{Khusnutdinova11} and the accuracy was controlled by monitoring conserved quantities. 

We note that the discretisation used for the terms $u_{ttx}$ and $w_{ttx}$ in \eqref{DNMcRBC2} is different to the one used in \cite{Khusnutdinova15}. The discretisation used in \cite{Khusnutdinova15} is not second order and therefore the scheme in that paper is not second order. The new scheme presented here used a second-order finite-difference approximation for these mixed derivatives to ensure that the scheme is second-order in time. This will lead to a small difference between the results in \cite{Khusnutdinova15, Khusnutdinova17} and the results generated using the new method presented here. In Section \ref{sec:Results} we show that the difference is negligible.

Let us denote the right-hand side of \eqref{DNMcRBDisc1} as $f_i^{(l)}$ and the right-hand side of \eqref{DNMcRBDisc2} as $g_i^{(l)}$, for the appropriate values of $i$, to simplify notation in the subsequent work. At this point we diverge from the method presented in \cite{Khusnutdinova15, Khusnutdinova17}. In each domain, defined by $l$, we have two ghost points: one at the left boundary and one at the right boundary. The exceptions are the first and $M$-th domain, where we do not have a ghost point on the left-hand side of the first domain and the right-hand side of the $M$-th domain. We apply the relations defined in \eqref{DNMcRBBCDisc} in these special domains to obtain a tridiagonal system. Before we write these matrices for each section, we rearrange at the boundaries between sections, for each layer, to obtain
\begin{align}
\tilde{f}^{(l)}_{0} &= f_{0}^{(l)} + 2 \epsilon u_{-1,j+1}^{(l)}, & \tilde{f}^{(l)}_{N_l} = f_{N_l}^{(l)} + 2 \epsilon u_{N_l+1,j+1}^{(l)}, \notag \\
\tilde{g}^{(l)}_{0} &= g_{0}^{(l)} + 2 \epsilon \beta_l w_{-1,j+1}^{(l)}, & \tilde{g}^{(l)}_{N_l} = g_{N_l}^{(l)} + 2 \epsilon \beta_l w_{N_l+1,j+1}^{(l)}. \label{DNMfgtilde}
\end{align}
We can now write \eqref{DNMcRBDisc1} in matrix form for each $l$, with two exceptional cases at the boundary. We will write the matrices for $w$, and the matrices for $u$ can be formed by setting $\alpha_{l} = \beta_{l} = c_{l} = 1$, $\gamma_{l} = - \delta_{l}$ and $f_{i} = g_{i}$. Therefore, in the first section, we have for $i=0,\dots,N_1$,
\begin{equation}
	\begin{pmatrix}
	 4 \epsilon \beta_{1} + h^2 & - 4 \epsilon \beta_{1} && \\[0.3em]
	- 2 \epsilon \beta_{1} &  4 \epsilon \beta_{1} + h^2 & - 2 \epsilon \beta_{1} & \\[0.3em]
	& \ddots &  \ddots  \\[0.3em]
	& - 2 \epsilon \beta_{1} &  4 \epsilon \beta_{1} + h^2 & - 2 \epsilon \beta_{1} \\[0.3em]
	&& - 2 \epsilon \beta_{1} &  4 \epsilon \beta_{1} + h^2
	\end{pmatrix}
	\begin{pmatrix}
	w_{0,j+1}^{(1)} \\[0.3em]
	w_{1,j+1}^{(1)} \\[0.3em]
	\vdots \\[0.3em]
	w_{N_1-1,j+1}^{(1)} \\[0.3em]
	w_{N_1,j+1}^{(1)}
	\end{pmatrix}
	=
	\begin{pmatrix}
	g_0^{(1)} \\[0.3em]
	g_1^{(1)} \\[0.3em]
	\vdots \\[0.3em]
	g_{N_1-1}^{(1)} \\[0.3em]
	\tilde{g}_{N_1}^{(1)}
	\end{pmatrix}.
\label{DNMcRBMatE1}
\end{equation}
Similarly for the final section we have, for $i = 0, \dots, N_{M}$,
\begin{equation}
	\begin{pmatrix}
	 4 \epsilon \beta_{M} + h^2 & - 2 \epsilon \beta_{M} && \\[0.3em]
	- 2 \epsilon \beta_{M} &  4 \epsilon \beta_{M} + h^2 & - 2\epsilon \beta_{M} & \\[0.3em]
	& \ddots &  \ddots  \\[0.3em]
	& - 2 \epsilon \beta_{M} &  4 \epsilon \beta_{M} + h^2 & - 2 \epsilon \beta_{M} \\[0.3em]
	&& - 4 \epsilon \beta_{M} &  4 \epsilon \beta_{M} + h^2
	\end{pmatrix}
	\begin{pmatrix}
	w_{0,j+1}^{(M)} \\[0.3em]
	w_{1,j+1}^{(M)} \\[0.3em]
	\vdots \\[0.3em]
	w_{N_M-1,j+1}^{(M)} \\[0.3em]
	w_{N_M,j+1}^{(M)}
	\end{pmatrix}
	=
	\begin{pmatrix}
	\tilde{g}_0^{(M)} \\[0.3em]
	g_{1}^{(M)} \\[0.3em]
	\vdots \\[0.3em]
	g_{N_M-1}^{(M)} \\[0.3em]
	g_{N_M}^{(M)}
	\end{pmatrix}.
\label{DNMcRBMatEM}
\end{equation}
Finally we have the generic matrix for all other regions, for $1 < l < M$, which takes the form
\begin{equation}
	\begin{pmatrix}
	 4 \epsilon \beta_{l} + h^2 & - 2 \epsilon \beta_{l} && \\[0.3em]
	- 2 \epsilon \beta_{l} &  4 \epsilon \beta_{l} + h^2 & - 2 \epsilon \beta_{l} & \\[0.3em]
	& \ddots &  \ddots  \\[0.3em]
	& - 2 \epsilon \beta_{l} &  4 \epsilon \beta_{l} + h^2 & - 2 \epsilon \beta_{l} \\[0.3em]
	&& - 2 \epsilon \beta_{l} &  4 \epsilon \beta_{l} + h^2
	\end{pmatrix}
	\begin{pmatrix}
	w_{0,j+1}^{(l)} \\[0.3em]
	w_{1,j+1}^{(l)} \\[0.3em]
	\vdots \\[0.3em]
	w_{N_l-1,j+1}^{(l)} \\[0.3em]
	w_{N_l,j+1}^{(l)}
	\end{pmatrix}
	=
	\begin{pmatrix}
	\tilde{g}_0^{(l)} \\[0.3em]
	g_{1}^{(l)} \\[0.3em]
	\vdots \\[0.3em]
	g_{N_l-1}^{(l)} \\[0.3em]
	\tilde{g}_{N_l}^{(l)}
	\end{pmatrix}.
\label{DNMcRBMatEl}
\end{equation}
We have now formed tridiagonal systems for the functions $u$ and $w$ and this system can be solved implicitly, in terms of the ghost points on the left boundary and right boundary, using the Thomas algorithm (e.g. \cite{Ames77}). Therefore we can express the solution at each point in terms of the explicit solution (if no ghost points exist) and a multiplicative factor of the ghost points. We denote this as
\begin{equation}
w_{i,j+1}^{(l)} = \phi_{i}^{(l)} + \psi_{i}^{(l)} w_{-1,j+1}^{(l)} + \omega_{i}^{(l)} w_{N_{l}+1,j+1}^{(l)},
\label{DNMcRBuImplicit}
\end{equation}
and we have a similar relationship for $u$. We note that $\psi_{i}^{(1)} = 0$ and $\omega_{i}^{(M)} = 0$ as there are no ghost points on these far boundaries. At this stage we note that the conditions on the interface do not contain any coupling terms; the coupling terms only appear at the current time step in the discretisation, in the functions $f$ and $g$. Therefore we can solve for the ghost points in each layer independently so we will only present the solution for $w$, and the solution for $u$ can be found in a similar way.

This implicit solution \eqref{DNMcRBuImplicit} is then substituted into \eqref{DNMcRBC1Disc} and \eqref{DNMcRBC2Disc2} to obtain a nonlinear system of equations. This is a complicated system and therefore we aim to simplify the expressions, if possible. Let us consider the matrices \eqref{DNMcRBMatE1} - \eqref{DNMcRBMatEl}, which were derived using \eqref{DNMcRBDisc1}. We can estimate the values of the coefficients $\psi_{i}^{(l)}$ and $\omega_{i}^{(l)}$ using the Thomas algorithm \cite{Ames77}. Explicitly following the steps of the Thomas algorithm to determine $w_{i, j + 1}^{(l)}$ from \eqref{DNMcRBMatEl} (or similarly $w_{i, j + 1}^{(1)}$, $w_{i, j + 1}^{(M)}$ from \eqref{DNMcRBMatE1}, \eqref{DNMcRBMatEM} respectively), we find that the coefficients $\psi_{i}^{(l)}$ and $\omega_{i}^{(l)}$ are dependent upon $h_l$ in a complicated way, leading to a continued fraction in terms of $h_l$. As $h_l$ is the step size, we are taking it to be reasonably small and therefore we can make an estimate of these coefficients by assuming $h_l \approx 0$. In this case we find that, for $i = 0, \dots, N_l$, we have
\begin{equation}
\psi_{i}^{(l)} = \frac{N_{l}+1-i}{N_{l}+2}, \quad \omega_{i}^{(l)} = \frac{1+i}{N_{l}+2}.
\label{DNMPsiOmega}
\end{equation}
Recalling that $N_{l}$ is the number of points in a given domain we see from \eqref{DNMPsiOmega} that, for a sufficiently large domain (or a sufficiently small value of $h_{l}$ to increase the number of points) we have that $\psi_{N_{l}}^{(l)} \approx 0$ and $\omega_{0}^{(l)} \approx 0$. Explicitly this means that the coefficient of the left ghost point in a given domain is approximately zero at the right boundary, and vice versa for the right ghost point at the left boundary. These have been calculated numerically for $N = 50,000$ and the value at the boundary was essentially zero $(\O{10^{-300}})$. Furthermore, it was found that the value falls below machine precision (i.e. $\O{10^{-16}}$) when $N = 500$, suggesting that a value of $h$ and the corresponding value of $N$ can be found for most domain sizes. This allows us to simplify the problem so that we only require the solution of two equations at each interface, in terms of the ghost points introduced at this interface.

In order to obtain the equations at the interface, we use \eqref{DNMcRBC1Disc} to express $w_{-1,j+1}^{(l+1)}$ in terms of $w_{N_l+1,j+1}^{(l)}$, making use of \eqref{DNMcRBuImplicit} to express $w_{N_l-1,j+1}^{(l)}$ and $w_{1,j+1}^{(l+1)}$ in terms of the ghost points. Substituting \eqref{DNMcRBuImplicit} into \eqref{DNMcRBC1Disc} gives
\begin{equation}
\phi_{N_l}^{(l)} + \omega_{N_l}^{(l)} w_{N_{l}+1,j+1}^{(l)} = \phi_{0}^{(l+1)} + \psi_{0}^{(l+1)} w_{-1,j+1}^{(l+1)}, 
\label{DNMC1PsiOmega}
\end{equation}
and therefore we have
\begin{equation}
w_{-1,j+1}^{(l+1)} = \frac{\phi_{N_l}^{(l)} - \phi_{0}^{(l+1)} + \omega_{N_l}^{(l)} w_{N_{l}+1,j+1}^{(l)}}{\psi_0^{(l+1)}}, \quad w_{1,j+1}^{(l+1)} = \phi_{1}^{(l+1)} + \frac{\psi_1^{(l+1)}}{\psi_0^{(l+1)}} \lb\phi_{N_l}^{(l)} - \phi_{0}^{(l+1)} + \omega_{N_l}^{(l)} w_{N_{l}+1,j+1}^{(l)} \rb.
\label{DNMC1Rearrange}
\end{equation}
Substituting \eqref{DNMcRBuImplicit} and \eqref{DNMC1Rearrange} into \eqref{DNMcRBC2Disc2} we obtain a quadratic equation in $w_{N_{l}+1,j+1}^{(l)}$ and therefore we need to solve a quadratic equation for each boundary. Explicitly we have
\begin{equation}
g_2 \lb w_{N_{l}+1,j+1}^{(l)} \rb^2 + g_1 w_{N_{l}+1,j+1}^{(l)} + g_0 = 0,
\label{DNMcRBuQuad}
\end{equation}
where we have
\begin{align}
g_2 &= -3 \kappa^2 \epsilon \alpha_{l} \lb \omega_{N_l-1}^{(l)} - 1 \rb^2 + 3 \frac{h_{l}^2}{h_{l+1}^2} \kappa^2 \epsilon \alpha_{l+1} \lb \frac{\omega_{N_{l}}^{(l)}}{\psi_{0}^{(l+1)}} \rb^2 \lb \psi_{1}^{(l+1)} - 1 \rb^2, \notag \\
g_1 &=  \lb 1 - \omega_{N_l-1}^{(l)} \rb \lb h_l \kappa^2 + 4 h_l \epsilon \beta_{l} + 6 \kappa^2 \epsilon \alpha_{l} \phi_{N_l-1}^{(l)} \rb + 6 \frac{h_{l}^2}{h_{l+1}^2} \kappa^2 \epsilon \alpha_{l+1} \lb \phi_{N_l}^{(l)} - \phi_0^{(l+1)} \rb \omega_{N_l}^{(l)} \lsq \frac{1 - \psi_1^{(l+1)}}{\psi_0^{(l+1)}} \rsq^2 \notag \\
&\quad + \lb 1 - \psi_{1}^{(l+1)} \rb \lb \frac{\omega_{N_l}^{(l)}}{\psi_0^{(l+1)}} \rb \lb \frac{h_{l}^2}{h_{l+1}} \rb \lb c_{l+1}^2 \kappa^2 + 4 \epsilon \beta_{l+1} - 6 \frac{\kappa^2}{h_{l+1}} \epsilon \alpha_{l+1} \phi_1^{(l+1)} \rb, \notag \\
g_0 &= 2 h_l \epsilon \beta_{l} \lb - 5 w_{N_l+1,j}^{(l)} + 5 w_{N_l-1,j}^{(l)} + 4 w_{N_l+1,j-1}^{(l)} - 4 w_{N_l-1,j-1}^{(l)} - w_{N_l+1,j-2}^{(l)} + w_{N_l-1,j-2}^{(l)} \rb \notag \\
&\quad -\phi_{N_l-1}^{(l)} \lb h_l \kappa^2 c_{l}^2  + 4 h_l \epsilon \beta_{l} \rb - \frac{h_{l}^{2}}{h_{l+1}} \lb \kappa^2 c_{l+1}^2 + 4 \frac{h_{l}^2}{h_{l+1}} \epsilon \beta_{l+1} \rb \lsq \phi_1^{(l+1)} + \frac{\phi_{N_{l}}^{(l)} - \phi_0^{(l+1)}}{\psi_0^{(l+1)}} \lb \psi_1^{(l+1)} - 1 \rb \rsq  \notag \\
&\quad + 3 \frac{h_{l}^2}{h_{l+1}^2} \kappa^2 \epsilon \alpha_{l+1} \lsq \lb \phi_1^{(l+1)} \rb^2 + \lb \frac{\phi_{N_l}^{(l)} - \phi_0^{(l+1)}}{\psi_0^{(l+1)}} \lb \psi_1^{(l+1)} - 1 \rb \rb^2 + 2 \lb \psi_1^{(l+1)} - 1 \rb \frac{\phi_1^{(l+1)} \lb \phi_{N_l}^{(l)} - \phi_0^{(l+1)} \rb}{\psi_0^{(l+1)}}  \rsq \notag \\
&\quad - 3 \kappa^2 \epsilon \alpha_{l} \lb \phi_{N_l-1}^{(l)} \rb^2 - 2 \frac{h_{l}^2}{h_{l+1}} \epsilon \beta_{l+1} \lb - 5 w_{1,j}^{(l+1)} + 5 w_{-1,j}^{(l+1)} + 4 w_{1,j-1}^{(l+1)} - 4 w_{-1,j-1}^{(l+1)} - w_{1,j-2}^{(l+1)} + w_{-1,j-2}^{(l+1)} \rb. \label{DNMcRBhcoef}
\end{align}
In the special case when $g_0 \equiv 0$ we can solve a linear equation on the boundary and numerical experiments have shown that, if the nonlinearity coefficient $\alpha_{l}$ is the same in both sections of the bar, this condition is approximately satisfied, so long as the ratio of the spatial discretisations in the sections is close to one. However, in the general case when $\alpha_{l}$ can change between sections (or indeed the spatial discretisation varies), we must choose the appropriate sign in the quadratic expression, and this is chosen to be consistent with the solution in the surrounding region.

Recall that we assumed $\psi_{N_{l}}^{(l)} \approx 0$ and $\omega_{0}^{(l)} \approx 0$ in each section. We now briefly comment on the method when these conditions do not hold. In contrast to the case outlined above, the conditions on the interface will provide two equations in terms of four variables; two ghost points at the current boundary and one ghost point from each of the adjacent sections. The exception to this is the first and last sections, where there would be three variables rather than four. 

To solve this system, one would solve the system between, say, the $(M - 1)$-th and $M$-th sections, where an implicit solution in terms of one of the ghost points would be obtained. This implicit solution is then used in the conditions on the interface between the $(M - 2)$-th and $(M - 1)$-th section, so that we would again be solving a system of two equation in three variables. This is repeated to the final section between the 1$^{\text{st}}$ and 2$^{\text{nd}}$ sections, where the system would be two equations for two ghost points. This can be solved explicitly and therefore the implicit solutions found previously become explicit. All cases considered herein satisfy the conditions $\psi_{N_{l}}^{(l)} \approx 0$ and $\omega_{0}^{(l)} \approx 0$.

\subsection{Extension to \texorpdfstring{$K$}{K} Layers}
\label{sec:cRBKLayers}
The model defined in \cite{Khusnutdinova09} could be extended to a bar with multiple layers, where the coupling occurs between two adjacent layers. The displacements in such a structure with $K$ layers are described by the equation system
\begin{align}
u_{tt}^{(l,1)} - c_{l,1}^2 u_{xx}^{(l,1)} &= 2 \epsilon \lsq -6 \alpha_{l,1} u_{x}^{(l,1)} u_{xx}^{(l,1)} + \beta_{l,1} u_{ttxx}^{(l,1)} - \delta_{l,1} \lb u^{(l,1)} - u^{(l,2)} \rb \rsq, \notag \\
u_{tt}^{(l,m)} - c_{l,m}^2 u_{xx}^{(l,m)} &= 2 \epsilon \lsq -6 \alpha_{l,m} u_{x}^{(l,m)} u_{xx}^{(l,m)} + \beta_{l,m} u_{ttxx}^{(l,m)} + \gamma_{l,m} \lb u^{(l,m-1)} - u^{(l,m)} \rb \right. \notag \\
&\left. \hspace{0.351\linewidth} -~ \delta_{l,m} \lb u^{(l,m)} - u^{(l,m+1)} \rb \rsq, \notag \\
u_{tt}^{(l,K)} - c_{l,K}^2 u_{xx}^{(l,K)} &= 2 \epsilon \lsq -6 \alpha_{l,K} u_{x}^{(l,K)} u_{xx}^{(l,K)} + \beta_{l,K} u_{ttxx}^{(l,K)} + \gamma_{l,K} \lb u^{(l,K-1)} - u^{(l,K)} \rb \rsq,
\label{cRBKLayer}
\end{align}
where we have $m = 2, \dots, K - 1$, and there are $l = 1, \dots, M$ sections in the structure. The subscript or superscript $l$ represents the section of the bar and similarly $m$ is an index for the layer. As before, the coefficients $\delta_{l,m}$ and $\gamma_{l,m}$ can be set to zero as necessary in the delaminated sections of the structure. The coefficient $\delta_{l,m}$ represents coupling with the layer below and $\gamma_{l,m}$ is coupling with the layer above the current layer. Note that we have introduced coefficients in all layers; these could be scaled to appear in only certain layers but, as we only consider the extension of the numerical scheme here, we do not require this scaling.

This system could be solved using the method outlined in Section \ref{sec:cRB} with a few modifications. When the discretisation is taken in the same way as before, using central difference approximations, the coupling terms will only appear at the current time step. Therefore, the matrices will take the same form as before i.e. \eqref{DNMcRBMatE1} - \eqref{DNMcRBMatEl} and the coupling terms are absorbed into the functions $f$. The conditions on the interface will have the same form with differing coefficients in the layers and therefore, as the functions $u^{(l,m)}$ can be written in terms of the ghost points in a particular layer, we can solve for the ghost points in each layer independently. The details are omitted for brevity, however the extension is natural. In theory this system could be extended further by relaxing conditions on the coefficients of the bonding layer, as outlined in \cite{Khusnutdinova09}, to obtain a more general long-wave model, where the bonding terms contain derivatives in $x$ or nonlinear terms. So long as the coupling does not contain derivatives in $t$, the method outlined above could be extended to accommodate such a generalisation.

\section{Numerical Results}
\label{sec:Results}
We use the numerical method described above to study the behaviour of strain solitary waves in a variety of layered elastic waveguides. Initially we will confirm the results of previous studies and then extend the analysis to the case of a finite delamination, as was done in \cite{Khusnutdinova17}, however we will analyse the effect of a finite delamination using the direct numerical method rather than the semi-analytical method. This restriction was enforced by the limitation of the direct numerical method, as outlined in Section \ref{sec:DNM3SeccRB}.

In all subsequent calculations we will take a strain solitary wave as the initial condition in the first section. To determine this initial condition, we find the strain solitary wave for the uncoupled equations by setting $\delta^{(l)} = \gamma^{(l)} = 0$ in \eqref{DNMcRBEq} and find the strain solitary wave from the resulting equation. We construct the initial condition for $w$ here and the initial condition for $u$ can be constructed in a similar way. If we consider the equation for $w$ in \eqref{DNMcRBEq}, setting $\gamma^{(l)} = 0$, the strain solitary wave takes the form (denoting $e(x,t) = w_{x}(x,t)$)
\begin{equation}
e(x, 0) = -\frac{\lb v^2 - c^2 \rb}{4 \alpha \epsilon} \sechn{2}{\frac{\sqrt{v^2 - c^2}}{2 v \sqrt{2 \epsilon \beta}} \lb x + x_0 \rb}, 
\label{StrainSol}
\end{equation} 
where $v$ can be thought of as the phase speed and $x_0$ is the phase shift. To obtain the initial condition for the scheme we integrate \eqref{StrainSol} with respect to $x$ to obtain
\begin{equation}
w(x, 0) = -\frac{v \sqrt{2 \epsilon \beta} \sqrt{v^2 - c^2}}{2 \alpha \epsilon} \lsq \tanh{\lb \frac{\sqrt{v^2 - c^2}}{2 v \sqrt{2 \epsilon \beta}} \lb x + x_0 \rb \rb} - 1 \rsq,
\label{DispIC}
\end{equation}
where the constant of integration is chosen so that the waves are propagating into an unperturbed medium. Therefore we have the initial condition for $w$ given by \eqref{DispIC} and, as we are taking the exact solitary wave solution in the first section of the bar, we could use the form of \eqref{DispIC} and calculate the solution at $w(x, \kappa)$ as
\begin{equation}
w(x, \kappa) = -\frac{v \sqrt{2 \epsilon \beta} \sqrt{v^2 - c^2}}{2 \alpha \epsilon} \lsq \tanh{\lb \frac{\sqrt{v^2 - c^2}}{2 v \sqrt{2 \epsilon \beta}} \lb x - v \kappa + x_0 \rb \rb} - 1 \rsq,
\label{DispIC2}
\end{equation}
where $v$ is as before. Similarly we could use a forward difference approximation in time for $w$ and we obtain the solution for $w(x,\kappa)$, and numerical simulations have shown that both methods agree. The parameter $v$ is chosen as $v = \sqrt{c^2 + 4 \alpha \epsilon A}$, so that we have amplitude $-A$ of the incident soliton.

\subsection{Confirmation of Previous Results}
\label{sec:PrevRes}
\begin{figure}[ht]
	\begin{center}
		\subfigure[Result at $t=1000$.]{\includegraphics[width=0.48\textwidth]{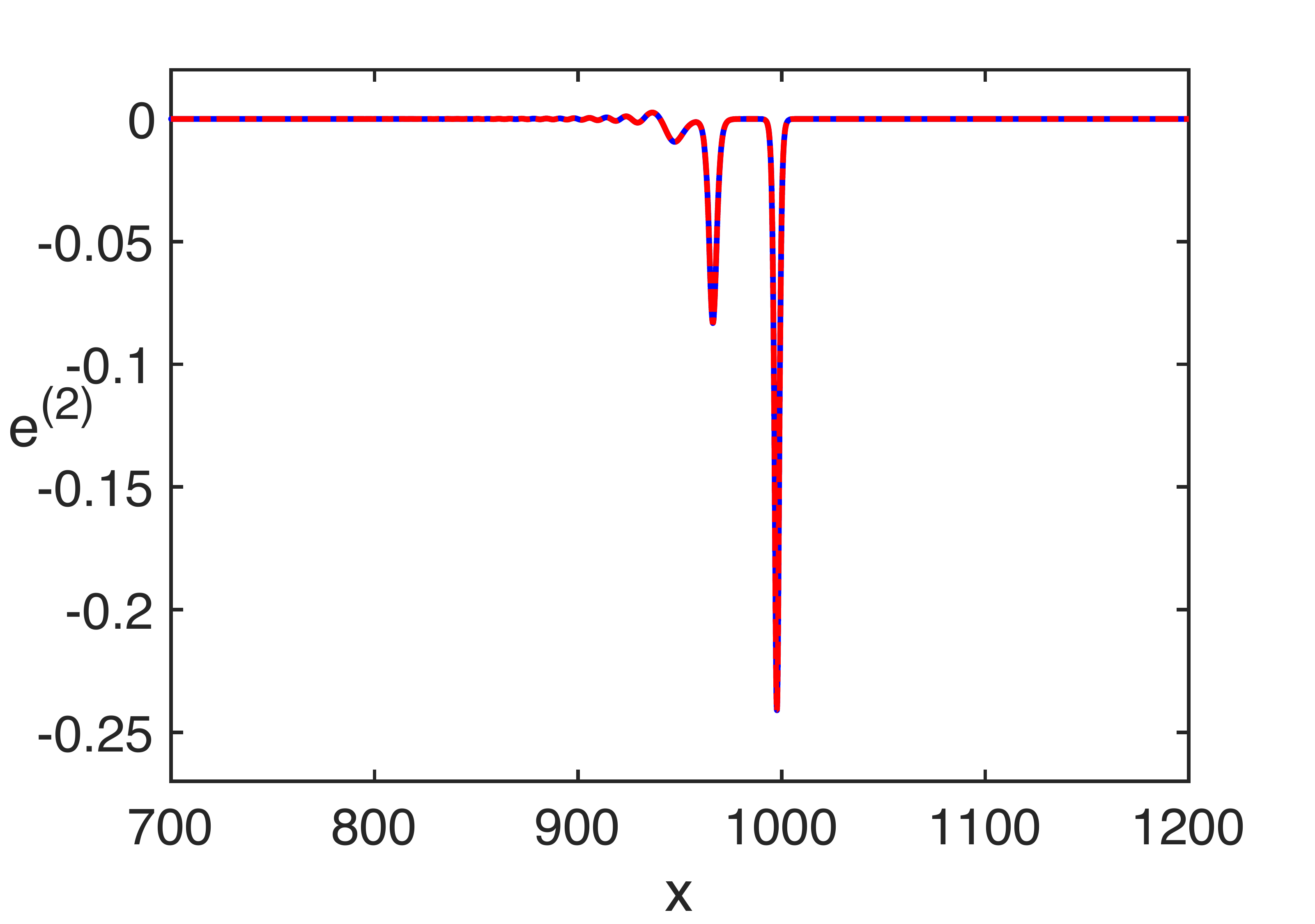}}
		\hspace{0.02\textwidth}
		\subfigure[Difference between the results.]{\includegraphics[width=0.48\textwidth]{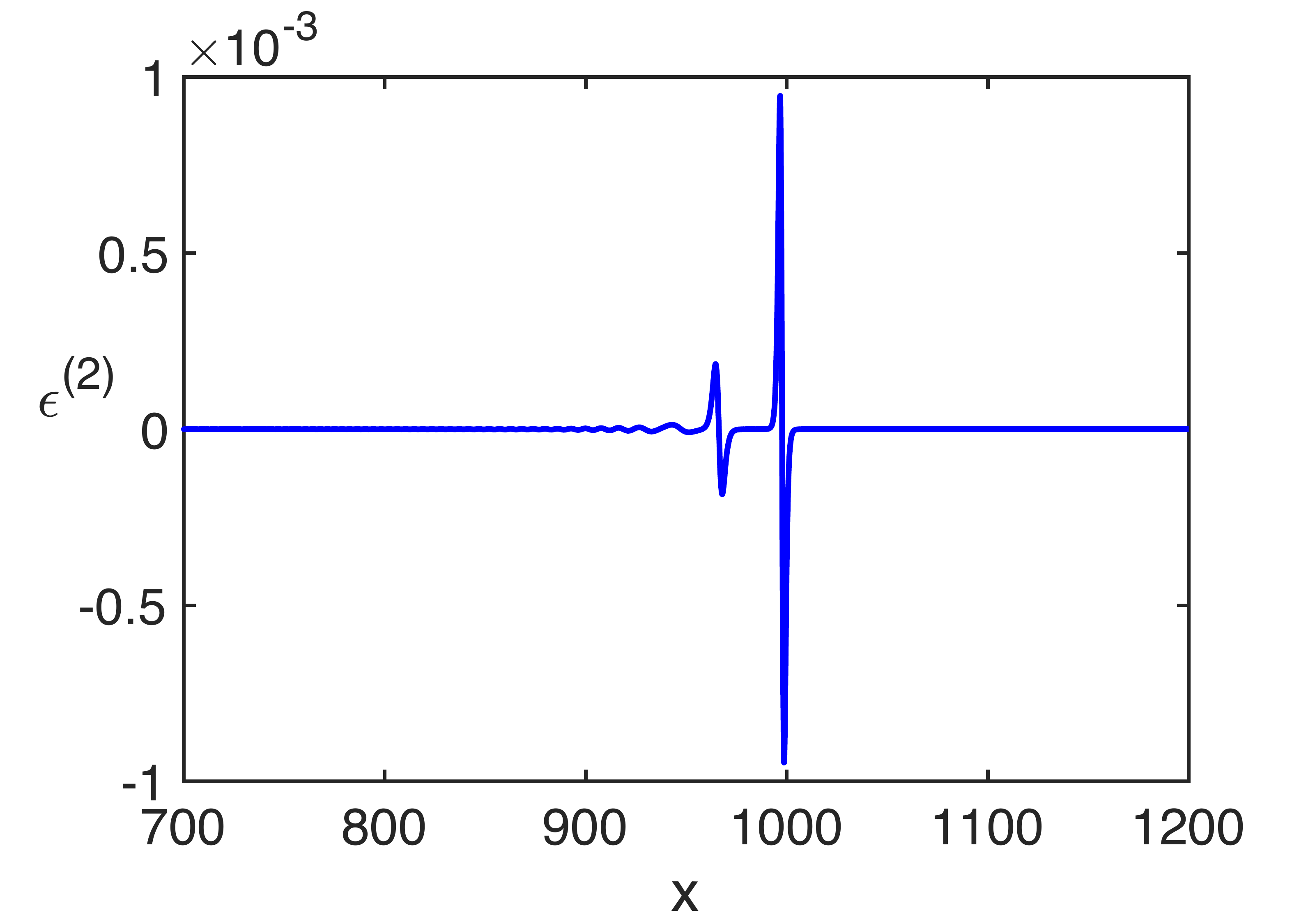}}
		\caption{Comparison of the results in \cite{Khusnutdinova15} for $n = 4$, $k = 2$, with amplitude $A = -0.175$ and $x_0 = -50$. Result is plotted in (a) for the new method (blue, solid line) and previous method (red, dashed line), and the difference between the methods is in (b).}
		\label{fig:DelamPBRes}
	\end{center}
\end{figure}

We now use the numerical method outlined in Section \ref{sec:DNM3SeccRB} to confirm the results obtained in \cite{Khusnutdinova15} and \cite{Khusnutdinova17}. In the case of a perfect bond, following \cite{Khusnutdinova15}, we solve \eqref{DNMcRBEq} with $c = \alpha = 1$ and $\beta$ takes the form
\begin{equation}
\beta(n,k) = \frac{n^2 + k^2}{n^2 \lb 1 + k^2 \rb},
\label{BetaVal}
\end{equation}
where $n$ is the number of layers in the bar and $k$ is defined by the geometry of the waveguide, namely for a cross-section of width $2a$ and layers of height $b$, we have $k = b/a$. We set $n = 4$ and $k = 2$ and, to correspond with Figure 9 in \cite{Khusnutdinova15}, we set $x_0 = -50$ and $A = -0.175$. The result is plotted at $t = 1000$ in Figure \ref{fig:DelamPBRes}a and the difference between the methods is plotted in Figure \ref{fig:DelamPBRes}b. We can see that there are clearly three solitons present, as predicted by the theory, and there is only a minor phase shift between the results, which can be accounted for by the corrected discretisation on the interface between sections. This difference is small as can be seen in Figure \ref{fig:DelamPBRes}b.

\begin{figure}[ht]
	\begin{center}
		\subfigure[Result in upper layer.]{\includegraphics[width=0.48\textwidth]{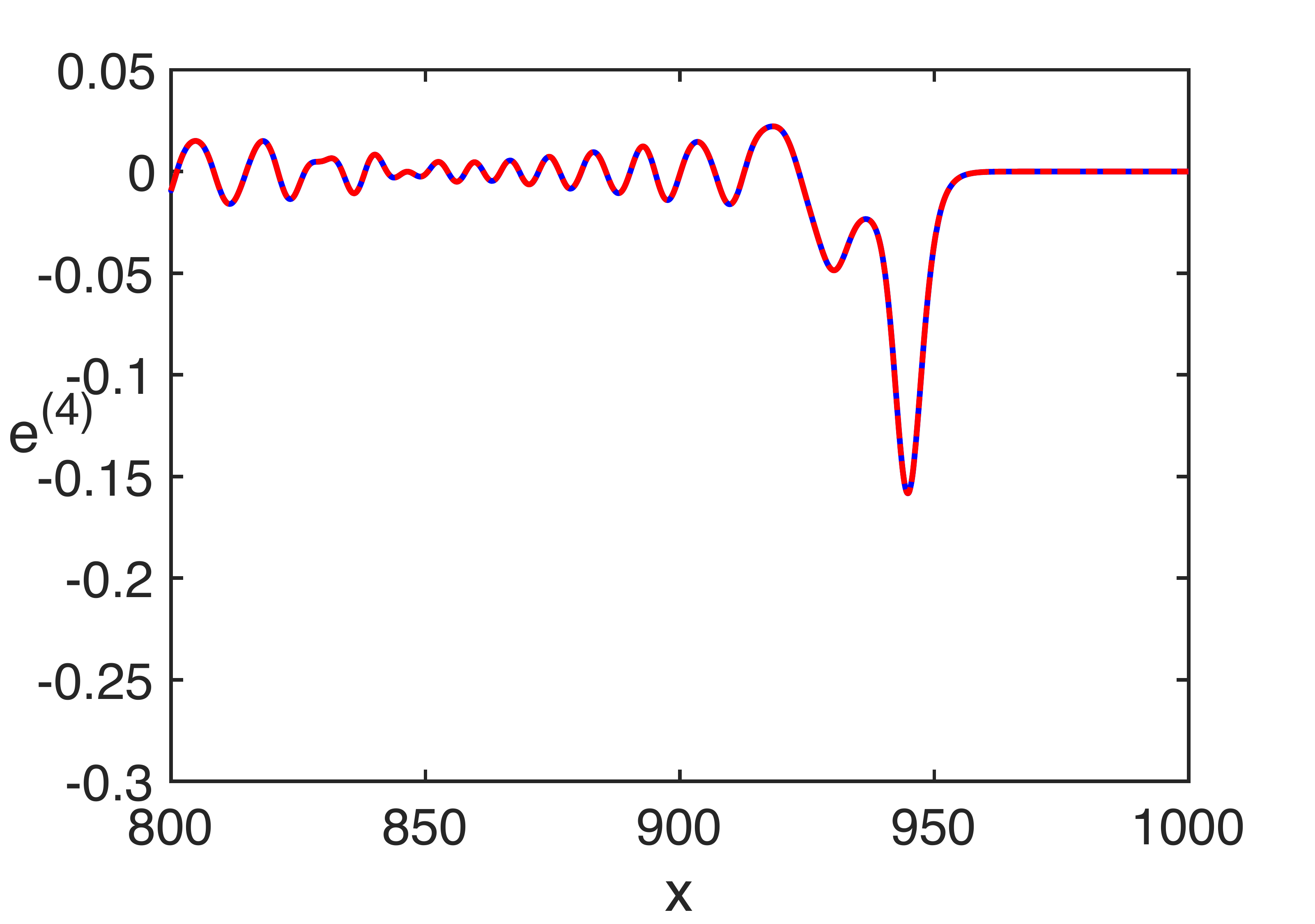}}
		\hspace{0.02\textwidth}
		\subfigure[Difference in upper layer.]{\includegraphics[width=0.48\textwidth]{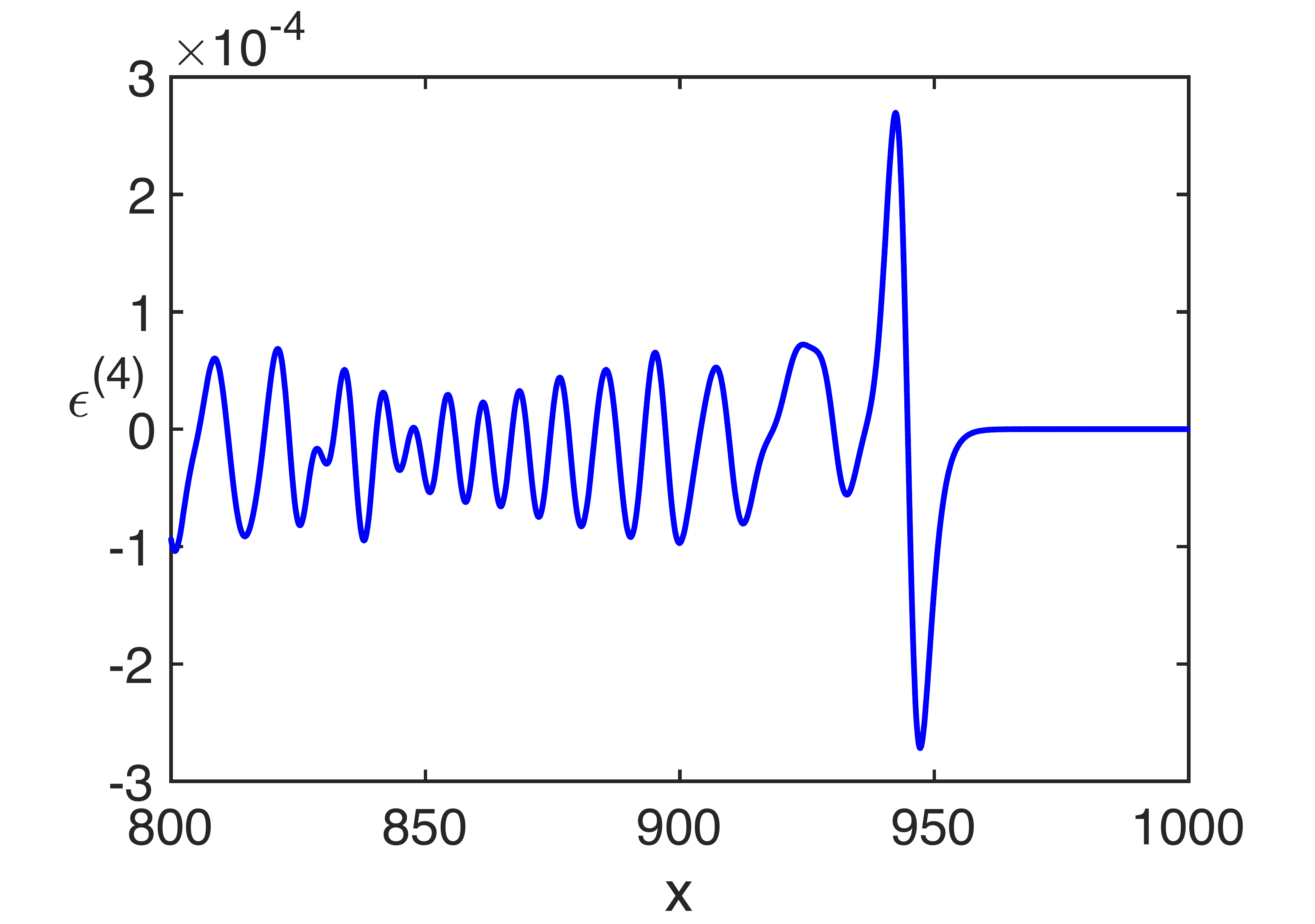}}
		\caption{Longitudinal strains in a two-layered bar with a soft bonding between the layers, in the upper layer, for a finite delamination. Result is plotted in (a) for the new method (blue, solid line) and previous method (red, dashed line), and the difference between the methods is in (b). Parameters are $\alpha = \beta = 1.05$, $c = 1.025$, $\delta = \gamma = 1/2$, $\epsilon = 0.05$ and we take $A = -0.25$.}
		\label{fig:cRBRes}
	\end{center}
\end{figure}

The results in \cite{Khusnutdinova17} are now checked for the case when we have a delamination of finite length, and we assume that there is a homogeneous section attached to the left of the bar (e.g. Figure 7 in \cite{Khusnutdinova17}). To correspond with the results in \cite{Khusnutdinova17}, we solve \eqref{DNMcRBEq} with the parameters $\alpha = \beta = 1.05$, $c = 1.025$, $\delta = \gamma = 1/2$, $\epsilon = 0.05$ and we take $A = -0.25$. The domain limits are set as $x_0 = -600$, $x_1 = -400$, $x_2 = 0$, $x_3 = 300$, $x_4 = 1000$. The agreement is similar in all sections and in both layers, so we only plot the results for the final bonded section and for the upper layer, and the difference between the schemes, in Figure \ref{fig:cRBRes}. We can see again that there is excellent agreement between the results and, as before, there is a minor phase shift present as identified by the difference.

\subsection{Finite Delamination in Perfectly Bonded Bar}
\label{sec:FinitePB}
An extension of the preliminary studies in \cite{Khusnutdinova15} can be performed for the case when there is a finite delamination, as can be seen in Figure \ref{fig:DelamBar3S}. This was done for the two-layered bar with a soft bonding in \cite{Khusnutdinova17}, however this will be discussed in the next section. We consider three sections so set $M = 3$. There are many parameters in this problem so, following Section \ref{sec:PrevRes}, we set $\alpha_{l} = c_{l} = 1$ in all sections, in the bonded regions ($l = 1, 3$) we take $\beta_{l} = 1$ and choose $\beta_{2}$ in the delaminated region in accordance with \eqref{BetaVal}.

\begin{figure}[ht]
	\begin{center}
		\subfigure[Section 2 at $t=500$.]{\includegraphics[width=0.48\textwidth]{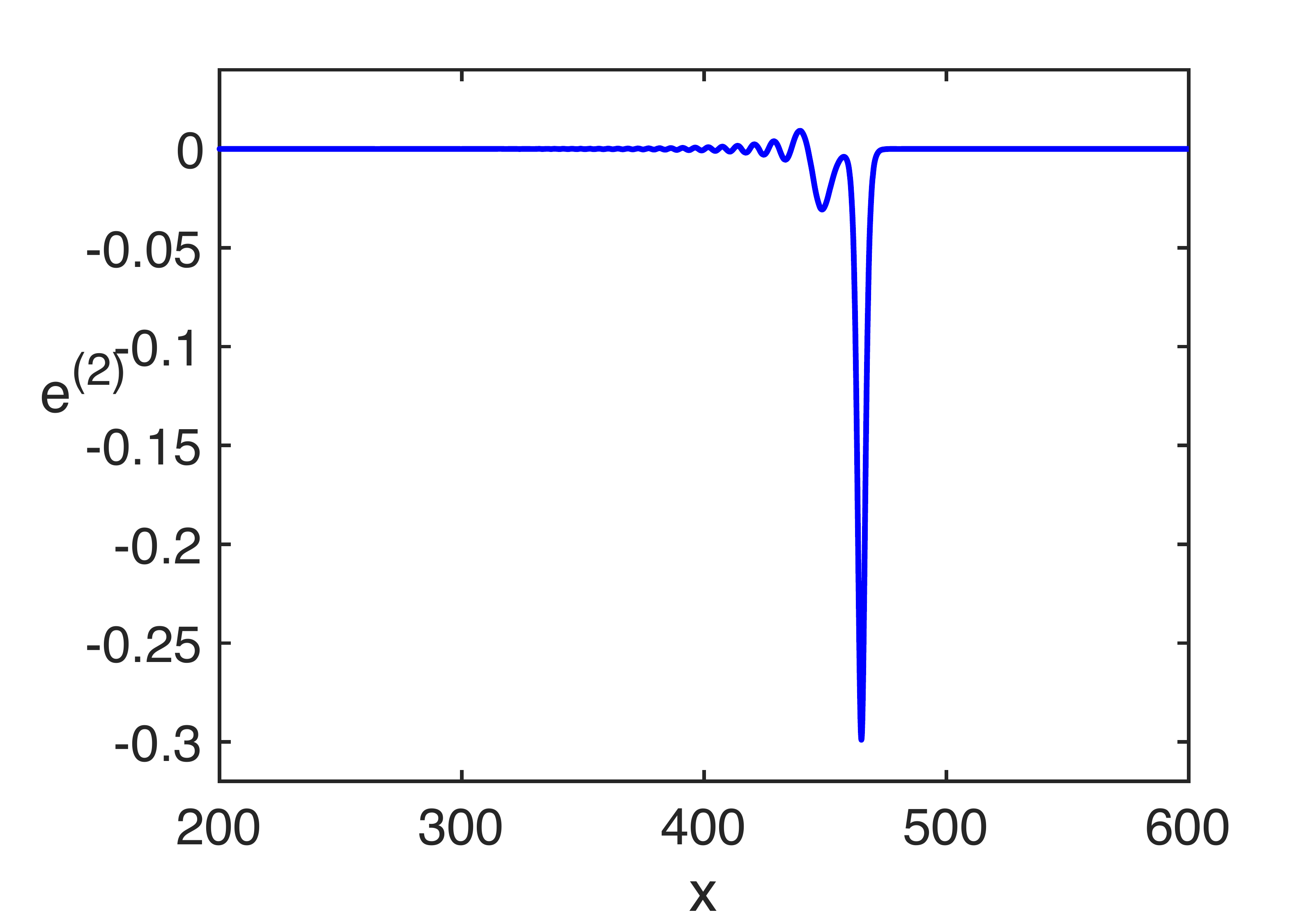}}
		\hspace{0.02\textwidth}
		\subfigure[Section 3 at $t=1200$.]{\includegraphics[width=0.48\textwidth]{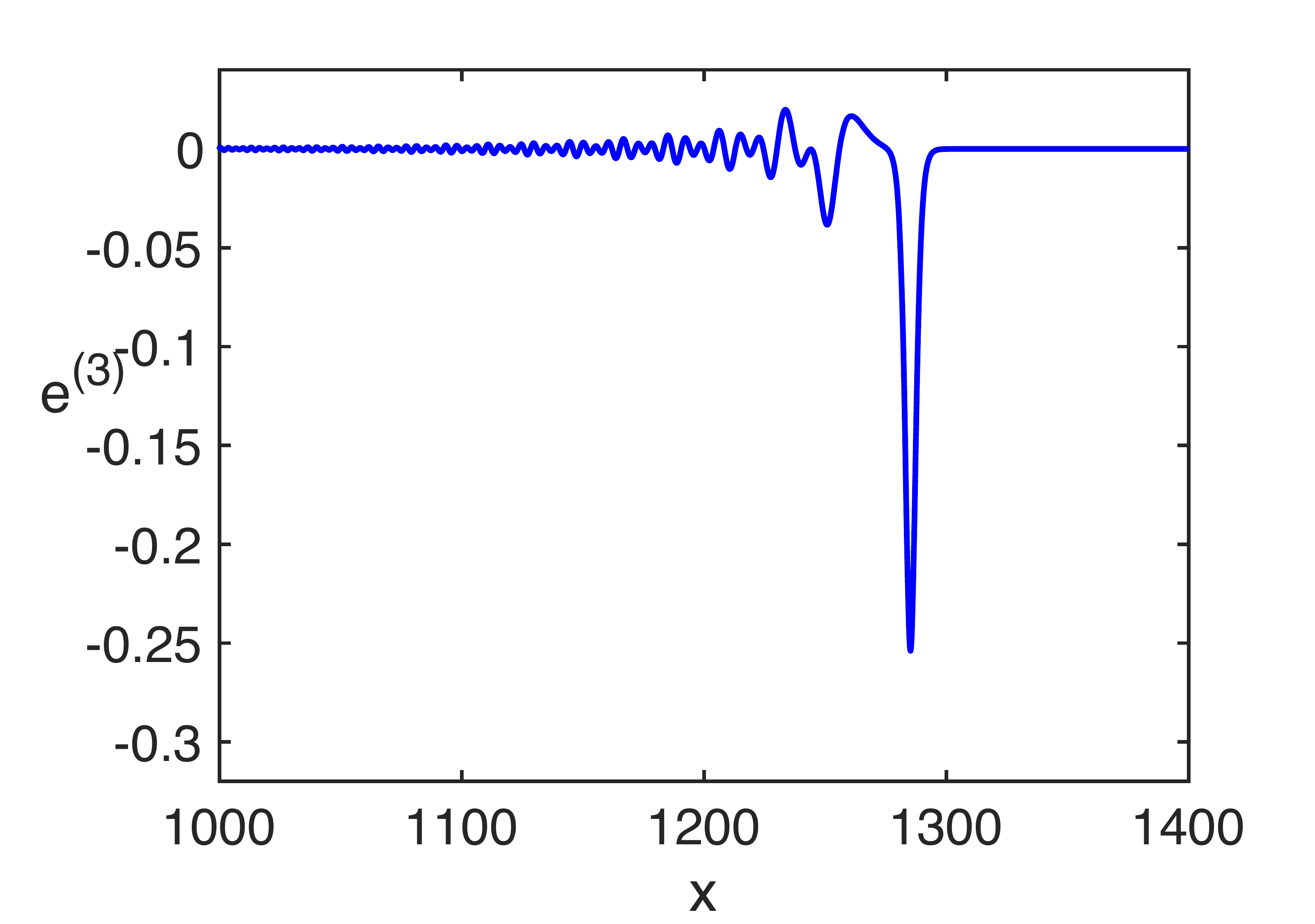}}
		\caption{The solution in each section of a perfectly bonded bar with $\alpha = c = 1$ in all sections, $\beta = 1$ in sections 1 and 3 and $\beta = 5/8$ in section 2, corresponding to $n = 2$ and $k = 2$, with initial position $x = -50$. The delamination length is $\ell = 300$ corresponding to 60 FWHM of the incident soliton i.e. $A \approx -0.2615$.}
		\label{fig:FDPBExample}
	\end{center}
\end{figure}

We want to study the effect of a finite delamination on the incident soliton, so we introduce the common measure Full Width at Half Magnitude (FWHM) of the incident soliton and the delamination length is calculated in terms of this measure. To determine a correspondence in the initial condition between $v$ and the FWHM of the soliton, using \eqref{StrainSol}, we solve
\begin{equation}
-\frac{\lb v^2 - c^2 \rb}{4 \alpha \epsilon} \sechn{2}{\frac{\sqrt{v^2 - c^2}}{2 v \sqrt{2 \epsilon \beta}} \frac{\mathrm{FWHM}}{2}} = -\frac{\lb v^2 - c^2 \rb}{8 \alpha \epsilon} \RA v = \sqrt{-\frac{c^2 \mathrm{FWHM}^2}{32 \beta \arccosh{\sqrt{2}}^2 - \mathrm{FWHM}^2}}.
\label{FWHMEq}
\end{equation}
For a FWHM of 5 this corresponds to $A \approx -0.2615$. The domain boundaries are set as $x_0 = -800$, $x_1 = 0$, $x_2 = \ell$, $x_3 = 1400$, where $\ell$ is the delamination length which we vary. The results are analysed at $t = 1200$ and compared to the expected result. As an illustrative example, we consider the case when $n = 2$ and $k = 2$ and plot the results in each section of the bar, as shown in Figure \ref{fig:FDPBExample}. We can see that the incident soliton fissions in the delaminated section and these solitons change their shape in the second bonded region. This poses the natural question: can we predict the amplitude of the lead soliton in the bonded region and can it be used to determine the delamination length?

Let us recall from \cite{Khusnutdinova08, Khusnutdinova15, Khusnutdinova17} that the transmitted waves in each region of a perfectly bonded bar can be described by KdV equations, and that theoretical estimates can be made for the amplitude of the lead soliton in the delaminated region of such a structure using the results of the Inverse Scattering Transform (IST), assuming that the solitons in each region are well separated (c.f. equation (4.4) in \cite{Khusnutdinova15} and (3.58) in \cite{Khusnutdinova17}). This can be extended to calculate the predicted amplitude of the lead soliton in the second bonded region, and in terms of our variables we have
\begin{equation}
A_{3} = A_{1} k_{2}^2 k_{3}^2, \quad k_{2} = \frac{1}{2} \lb \sqrt{1 + \frac{8}{\beta_{2}}} - 1 \rb, \quad k_{3} = \frac{1}{2} \lb \sqrt{1 + 8 \beta_{2}} - 1 \rb,
\label{AmpPrediction}
\end{equation}
where $A_{1}$ is the amplitude of the incident soliton i.e. $A_{1} = A$, $A_{3}$ is the amplitude of the lead soliton in the second bonded region, and $k_{2}$, $k_{3}$ are the eigenvalues corresponding to the lead soliton amplitude, as determined by the IST. It follows that, as the length of the delaminated region is reduced, the amplitude of the lead soliton in the third region will tend towards the initial amplitude, $A_{1}$. Denoting the calculated numerical solution as $A_{\text{num}}$, we introduce a measure of the amplitude of the lead soliton in the third section of the bar in comparison to the incident soliton as
\begin{equation}
\sigma = \frac{A_{\text{num}} - A_{1}}{A_{3} - A_{1}} \times 100.
\label{SolAmpSigma}
\end{equation}
We note that the prediction is based upon the leading order weakly nonlinear solution, therefore the actual amplitude will be slightly different and this could be accounted for by higher order corrections, which are beyond the scope of this study. 

\begin{figure}[ht]
	\begin{center}
		\subfigure[Varying $\beta$, $\epsilon = 0.10$]{\includegraphics[width=0.45\linewidth, trim = 0mm 0mm 0mm 0mm]{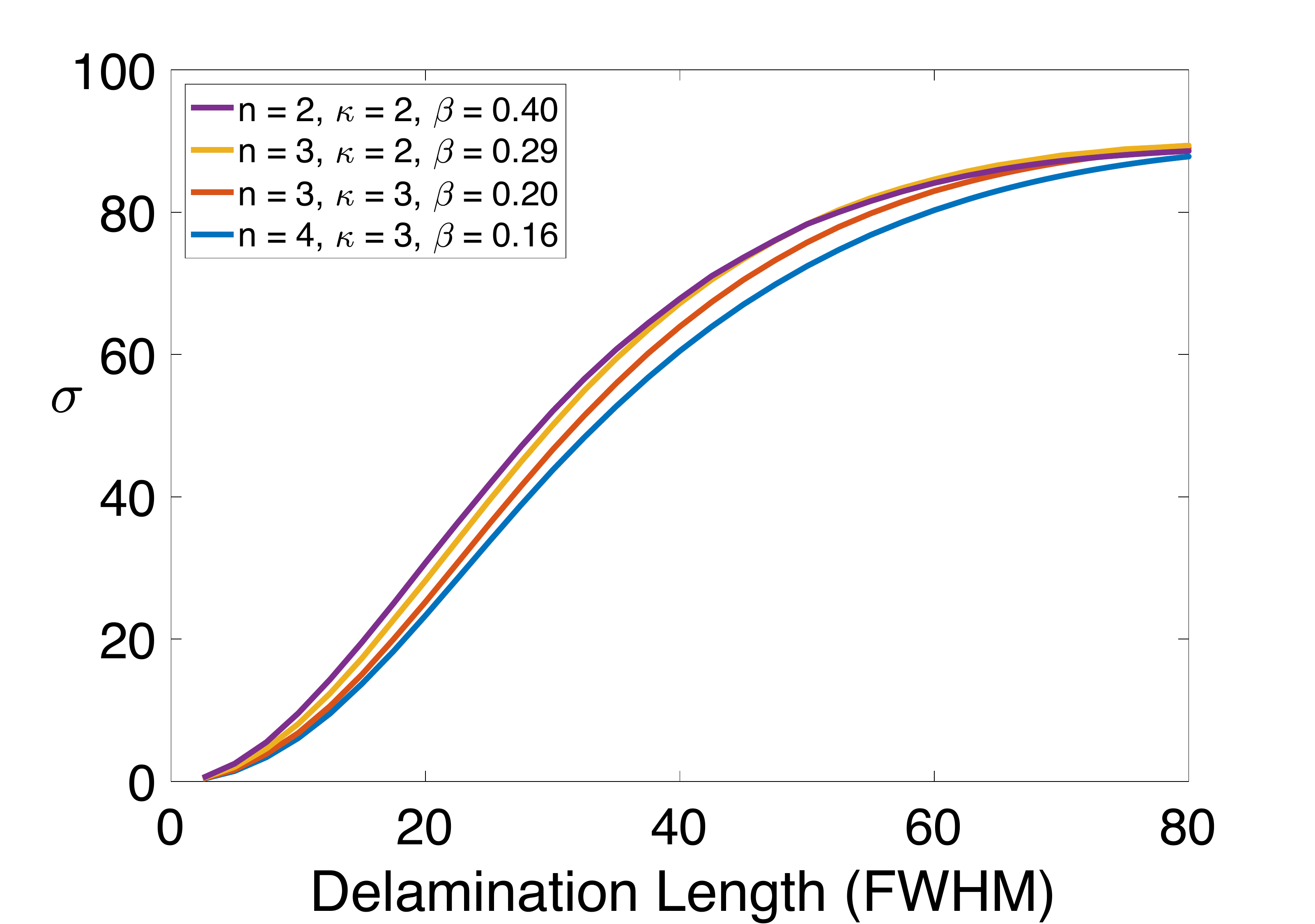}}
		\hspace{0.05\linewidth}
		\subfigure[Varying $\epsilon$, $\beta = 0.16$]{\includegraphics[width=0.45\linewidth, trim = 0mm 0mm 0mm 0mm]{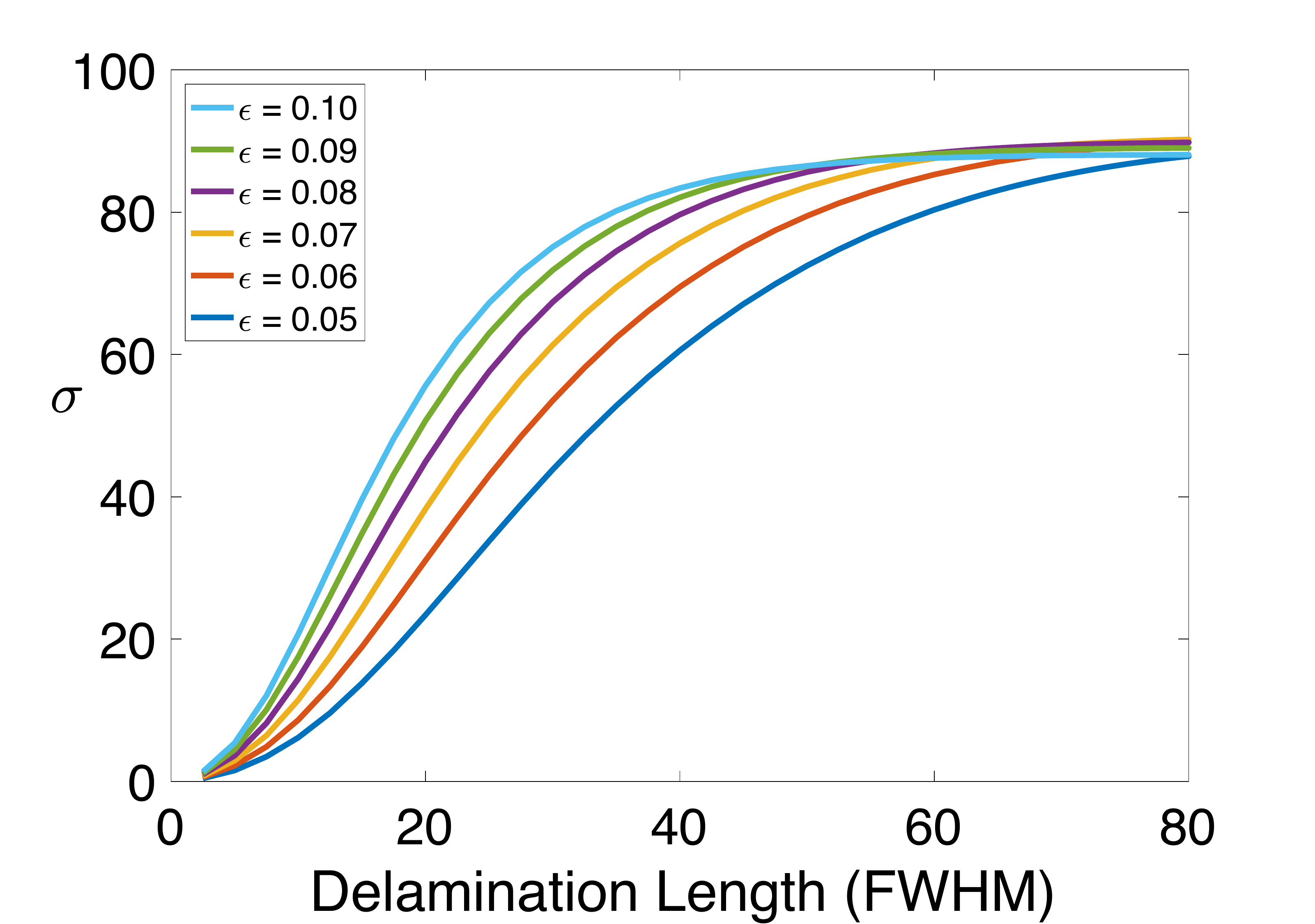}}
		\caption{Graphs of the change in amplitude of the lead transmitted soliton in comparison to the incident soliton, as measured by $\sigma$. Graph (a) corresponds to changing values of $\beta$ (changing the geometry of the waveguide), while graph (b) corresponds to changing values of $\epsilon$.}
		\label{fig:FDPB}
	\end{center}
\end{figure}

We consider two cases: the effect of varying $\beta$ for fixed $\epsilon$, and varying $\epsilon$ for fixed $\beta$. In each case we assume $\ell$ varies from $\ell = 0$ to $\ell = 400$ in steps of $2.5\times \text{FWHM}$. We denote the value of $\ell$ in terms of the FWHM of the incident soliton in Figures \ref{fig:FDPB} and \ref{fig:FDcRB}.

For the case of varying $\beta$, we take $n = 2,3,4$ and $k = 1,2,3$ and analyse the value of $\sigma$ at various delamination lengths, for an appropriate range of $\beta$ values in Figure \ref{fig:FDPB}a. We can see that, as $\beta$ decreases, for a fixed value of $\sigma$ the corresponding value of $\ell$ increases. Furthermore, for a given structure the value of $\beta$ is known and, if a value of $\sigma$ can be measured in the bonded region, the corresponding delamination length, $\ell$, can be found. The effect of varying $\epsilon$ instead of $\beta$ is shown in Figure \ref{fig:FDPB}b. Similarly to the previous case as $\epsilon$ decreases, for a fixed value of $\sigma$, the corresponding value of $\ell$ increases and we can determine the delamination length for a known value of $\beta$ and $\epsilon$ for a measured value of $\sigma$.

\subsection{Finite Delamination in Bar with Soft Bonding}
\label{sec:FiniteSB}
\begin{figure}[ht]
	\begin{center}
		\includegraphics[width=0.48\linewidth, trim = 0mm 0mm 0mm 0mm]{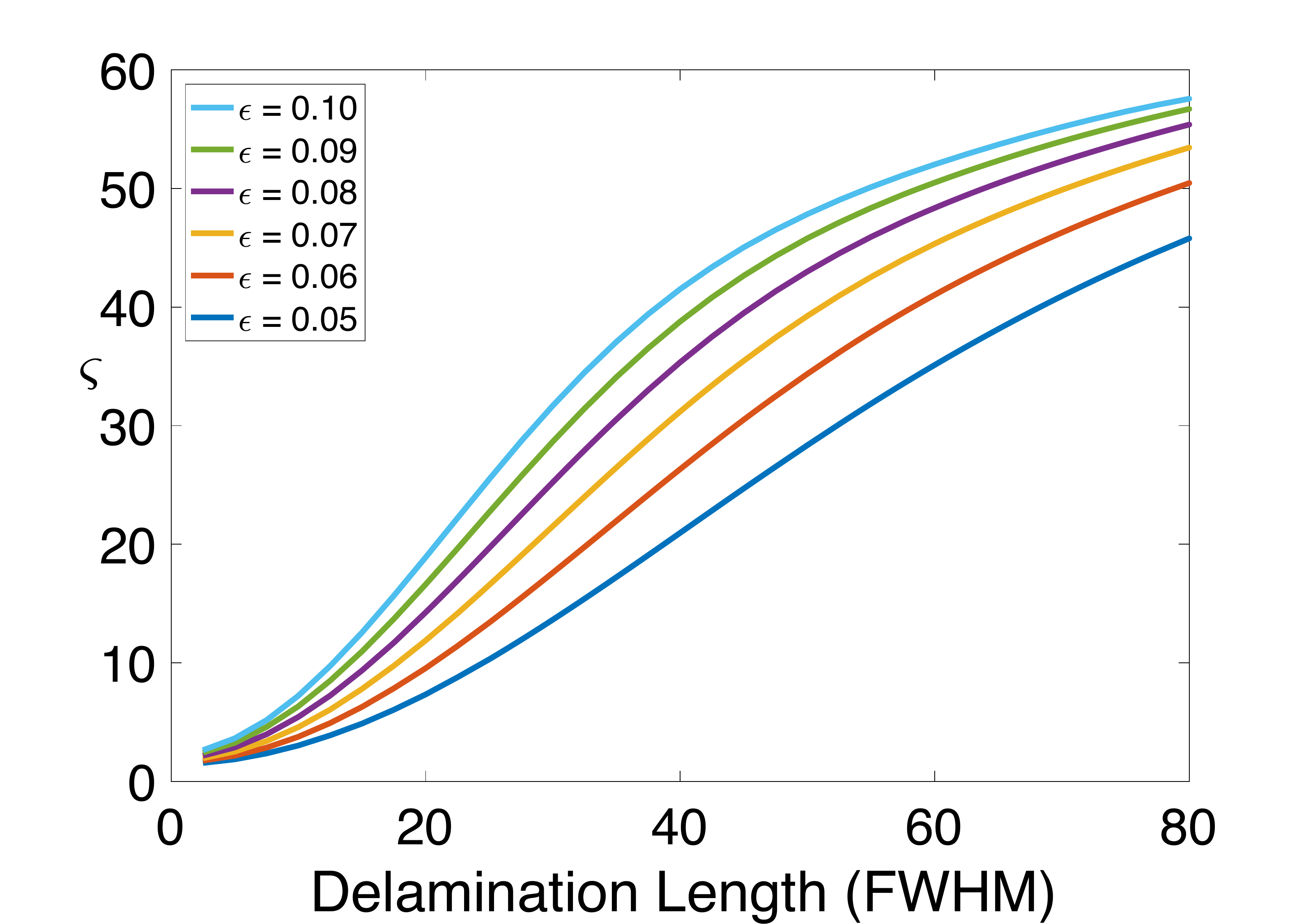}
		\caption{Graph of the change in amplitude of the main wave structure in the second bonded region, in comparison to the radiating solitary wave in the first bonded region, as measured by $\varsigma$, for various values of $\epsilon$.}
		\label{fig:FDcRB}
	\end{center}
\end{figure}

Let us consider the case of a finite delamination in a bar with a soft bond between the layers. This was previously studied in \cite{Khusnutdinova17} but the results were for the semi-analytical numerical technique only, due to the limitations of the direct numerical method. We now compute the results using our improved numerical method and compare to the previous results in \cite{Khusnutdinova17}, as well as investigating the effect of varying $\epsilon$. In contrast to the previous case for a perfectly bonded bar, the leading order weakly nonlinear solution is governed by coupled Ostrovsky equations in the bonded regions and therefore we cannot use the IST to predict the amplitude of the waves in the second bonded region. Therefore, following \cite{Khusnutdinova17}, we denote the amplitude of the soliton or wave packet in each region as $A_{l}$, where $l$ is the region, and therefore we have the measure
\begin{equation}
\varsigma = \frac{\abs{A_{4} - A_{2}}}{A_{2}} \times 100,
\label{cRBAmpEq}
\end{equation}
where $A_{2}$ is the amplitude of the radiating solitary wave in the first bonded region and $A_{4}$ is the amplitude of the main wave structure in the second bonded region. The results are plotted for various values of $\epsilon$ in Figure \ref{fig:FDcRB}. The same result that was observed for the perfectly bonded bar applies here i.e. as the value of $\epsilon$ is decreased, the value of $\ell$ increases by the same factor, for a fixed value of amplitude decrease.

The results in Figure \ref{fig:FDcRB} can be compared to the corresponding results for the semi-analytical method in \cite{Khusnutdinova17}, which are presented for $\epsilon = 0.05$ and therefore we compare to the corresponding line in Figure \ref{fig:FDcRB} (the lowest line). These results are in agreement and this suggests that the results for the semi-analytical method are comparable to the direct numerical method, as before. Here we have extended the results from \cite{Khusnutdinova17} to show the range of results for different values of $\epsilon$ and a larger delamination length.

\section{Conclusions}
\label{sec:Conclusions}
In this paper we have developed an improved numerical technique for the system of Boussinesq-type equations describing long longitudinal waves in a layered elastic bar with delamination. This system consists of uncoupled or coupled Boussinesq equations in each section of the bar with continuity conditions on the interface. A numerical technique for such systems was introduced previously in \cite{Khusnutdinova15} and extended in \cite{Khusnutdinova17}, however this method was limited to only two sections at a time. The work in this paper has introduced an improved technique that allows for the solution to be computed in all sections of a bar at the same time, for any number of sections and layers.

This improved numerical technique was used to compute the solutions observed in the previous studies, to confirm its validity. The results are in excellent agreement with a minor phase shift and confirm that the numerical technique is generating consistent results. We then computed the results for a finite delamination in a bar with a soft bonding and again found excellent agreement with the previous results. The study was extended to the case when the value of $\epsilon$ varies and we showed that the length of the delamination could be determined from the changes to the incident soliton, specifically the amplitude of the leading wave packet. The same approach was used for the perfectly bonded bar and results were determined for varying $\beta$ and $\epsilon$, where the delamination length can again be determined from the amplitude changes.

This improvement addressed the limitations of the previous model and allows for the modelling of a wider range of configurations of the bar. Furthermore, it allows for the reflected waves to be studied as well as the transmitted waves in each section of the bar, a clear improvement on previous direct numerical methods. It was also shown that the method can be extended to the case with multiple layers and different coupling terms, as previously theorised in \cite{Khusnutdinova09}. These extensions lie beyond the scope of this study.

\section*{Acknowledgements}
The author would like to acknowledge that the type of problem solved by this numerical method originated from a PhD project, funded by an EPSRC bursary, devoted to the modelling of nonlinear waves in layered elastic waveguides with delamination under the supervision of K. R. Khusnutdinova. The author would like to thank the supervisor for posing the problems discussed herein.

\section*{References}
\bibliographystyle{elsarticle-num}
\bibliography{Research} 

\begin{thebibliography}{10}
\expandafter\ifx\csname url\endcsname\relax
  \def\url#1{\texttt{#1}}\fi
\expandafter\ifx\csname urlprefix\endcsname\relax\def\urlprefix{URL }\fi
\expandafter\ifx\csname href\endcsname\relax
  \def\href#1#2{#2} \def\path#1{#1}\fi

\bibitem{Samsonov01}
A.~M. Samsonov, Strain Solitons in Solids and How to Construct Them, CRC Press,
  2001.

\bibitem{Porubov03}
A.~V. Porubov, Amplification of Nonlinear Strain Waves in Solids, World
  Scientific, 2003.

\bibitem{Dreiden88}
G.~V. Dreiden, Y.~I. Ostrovsky, A.~M. Samsonov, I.~V. Semenova, E.~V.
  Sokurinskaya, Formation and propagation of deformation solitons in a
  nonlinearly elastic solid, Zh. Tekh. Fiz. 58 (1988) 2040--2047.

\bibitem{Samsonov98}
A.~M. Samsonov, G.~V. Dreiden, A.~V. Porubov, I.~V. Semenova,
  Longitudinal-strain soliton focusing in a narrowing nonlinearly elastic rod,
  Phys. Rev. B 57 (1998) 5778.

\bibitem{Semenova05}
I.~V. Semenova, G.~V. Dreiden, A.~M. Samsonov, On nonlinear wave dissipation in
  polymers, Proc. SPIE, Optical Diagnostics 5880 (2005) 588006.

\bibitem{Khusnutdinova08}
K.~R. Khusnutdinova, A.~M. Samsonov, Fission of a longitudinal strain solitary
  wave in a delaminated bar, Phys. Rev. E 77 (2008) 066603.

\bibitem{Khusnutdinova09}
K.~R. Khusnutdinova, A.~M. Samsonov, A.~S. Zakharov, Nonlinear layered lattice
  model and generalized solitary waves in imperfectly bonded structures, Phys.
  Rev. E 79 (2009) 056606.

\bibitem{Dreiden08}
G.~V. Dreiden, K.~R. Khusnutdinova, A.~M. Samsonov, I.~V. Semenova, Comparison
  of the effect of cyanoacrylate- and polyurethane-based adhesives on a
  longitudinal strain solitary wave in layered polymethylmethacrylate
  waveguides, J. Appl. Phys. 104 (2008) 086106.

\bibitem{Dreiden11}
G.~V. Dreiden, A.~M. Samsonov, I.~V. Semenova, K.~R. Khusnutdinova, Observation
  of a radiating bulk strain soliton in a solid-state waveguide, Tech. Phys. 56
  (2011) 889--892.

\bibitem{Dreiden12}
G.~V. Dreiden, K.~R. Khusnutdinova, A.~M. Samsonov, I.~V. Semenova, Bulk strain
  solitary waves in bonded layered polymeric bars with delamination, J. Appl.
  Phys. 112 (2012) 063516.

\bibitem{Dreiden14a}
G.~V. Dreiden, A.~M. Samsonov, I.~V. Semenova, Observation of bulk strain
  solitons in layered bars of different materials, Tech. Phys. Lett. 40 (2014)
  1140--1141.

\bibitem{Khusnutdinova11}
K.~R. Khusnutdinova, K.~R. Moore, Initial-value problem for coupled
  {B}oussinesq equations and a hierarchy of {O}strovsky equations, Wave Motion
  48 (2011) 738--752.

\bibitem{Grimshaw17}
R.~H.~J. Grimshaw, K.~R. Khusnutdinova, K.~R. Moore, Radiating solitary waves
  in coupled {B}oussinesq equations, IMA J. Appl. Math. 82 (2017) 802--820.

\bibitem{Gerald84}
C.~F. Gerald, P.~O. Wheatley, Applied Numerical Analysis, World Student Series,
  Addison-Wesley, 1984.

\bibitem{Smith85}
G.~D. Smith, Numerical Solution of Partial Differential Equations: Finite
  Difference Methods, Oxford Applied Mathematics, Clarendon Press, 1985.

\bibitem{Drazin89}
P.~G. Drazin, R.~S. Johnson, Solitons: An Introduction, Cambridge University
  Press, 1989.

\bibitem{Marchant02}
T.~R. Marchant, N.~F. Smyth, The initial boundary value problem for the {KdV}
  equation on the negative quarter-plane, Proc. R. Soc. Lond. A 458 (2002)
  857--871.

\bibitem{Bratsos98}
A.~G. Bratsos, E.~H. Twizell, An explicit finite-difference scheme for the
  solution of the {K}adomtsev-{P}etviashvili equation, Int. J. Comput. Math. 68
  (1998) 175--187.

\bibitem{Khusnutdinova15}
K.~R. Khusnutdinova, M.~R. Tranter, Modelling of nonlinear wave scattering in a
  delaminated elastic bar, P. Roy. Soc. A 471~(2183) (2015) 20150584.

\bibitem{Khusnutdinova17}
K.~R. Khusnutdinova, M.~R. Tranter, On radiating solitary waves in bi-layers
  with delamination and coupled {O}strovsky equations, Chaos 27 (2017) 013112.

\bibitem{Khusnutdinova17a}
K.~R. Khusnutdinova, M.~R. Tranter, Scattering of bulk strain solitary waves in
  bi-layers with delamination, Procedia Eng. 199 (2017) 1533--1538.

\bibitem{Ames77}
W.~F. Ames, Numerical Methods for Partial Differential Equations, 2nd Edition,
  Academic Press, Inc., 1977.

\end{thebibliography}

\end{document}